\documentclass[12pt]{iopart}
\usepackage{graphicx}
\begin{document}
\hyphenation{fer-ro-mag-net fer-ro-mag-nets meta-sta-ble meta-sta-bi-lity
con-fi-gu-ra-tion con-fi-gu-ra-tions expo-nen-tially}

\newcommand{\beq}{\begin{equation}}
\newcommand{\eeq}{\end{equation}}
\newcommand{\bea}{\begin{eqnarray}}
\newcommand{\eea}{\end{eqnarray}}

\newcommand{\abs}[1]{\vert#1\vert}
\newcommand{\ap}{{\rm ap}}
\newcommand{\bin}[2]{{#1\choose#2}}
\newcommand{\bl}{\circ{\hskip .21mm}}
\renewcommand{\c}{{\rm c}}
\newcommand{\ca}{{\rm S}}
\newcommand{\co}{{\rm co}}
\newcommand{\conf}{{\rm conf}}
\renewcommand{\d}{{\rm d}}
\newcommand{\del}{\delta}
\newcommand{\diag}{{\rm diag}}
\newcommand{\dmean}[1]{\langle#1\rangle}
\newcommand{\dpar}{\partial}
\newcommand{\dyn}{{\rm dyn}}
\renewcommand{\e}{{\rm e}}
\newcommand{\eps}{\varepsilon}
\newcommand{\eq}{{\rm eq}}
\newcommand{\frad}[2]{\displaystyle{\displaystyle#1\over\displaystyle#2}}
\newcommand{\g}{\gamma}
\newcommand{\he}{{\rm H}}
\renewcommand{\i}{{\rm i}}
\newcommand{\im}{{\mathop{\rm Im\ }}}
\newcommand{\lam}{\lambda}
\newcommand{\lra}{\Longleftrightarrow}
\renewcommand{\max}{{\rm max}}
\newcommand{\mean}[1]{\langle#1\rangle}
\newcommand{\no}{\bullet{\hskip .21mm}}
\newcommand{\prob}{{\rm Prob}}
\newcommand{\s}{\sigma}
\renewcommand{\sp}{{\rm sp}}
\renewcommand{\t}{\tau}
\newcommand{\var}{\mathop{\rm Var}}
\newcommand{\vecn}{{\bf n}}
\newcommand{\vecz}{{\bf 0}}
\newcommand{\w}{\omega}
\newcommand{\C}{{\cal C}}
\newcommand{\En}{{\cal E}}
\newcommand{\F}{{\cal F}}
\renewcommand{\H}{{\cal H}}
\newcommand{\I}{{\cal I}}
\newcommand{\HB}{{\rm HB}}
\newcommand{\M}{{\rm M}}
\newcommand{\N}{{\cal N}}
\renewcommand{\S}{\Sigma}
\newcommand{\T}{{\cal T}}
\eqnobysec

\title[Metastability in zero-temperature dynamics]
{Metastability in zero-temperature dynamics: Statistics of attractors}
\author{C Godr\`eche\dag\ and J M Luck\ddag}

\address{\dag\ Service de Physique de l'\'Etat Condens\'e,
CEA Saclay, 91191~Gif-sur-Yvette cedex, France}

\address{\ddag\ Service de Physique Th\'eorique\footnote{URA 2306 of CNRS},
CEA Saclay, 91191~Gif-sur-Yvette cedex, France}

\begin{abstract}
The zero-temperature dynamics of simple models such as Ising ferro\-magnets
provides, as an alternative to the mean-field situation,
interesting examples of dynamical systems with many attractors
(absorbing configurations, blocked confi\-gurations,
zero-temperature metastable states).
After a brief review of metastability in the mean-field ferromagnet
and of the droplet picture,
we focus our attention onto zero-temperature single-spin-flip dynamics
of ferromagnetic Ising models.
The situations leading to metastability are characterized.
The statistics and the spatial structure of the at\-trac\-tors thus obtained
are investigated, and put in perspective with uniform a priori ensembles.
We review the vast amount of exact results available in one dimension,
and present original results on the square and honeycomb lattices.
\end{abstract}

\eads{\mailto{godreche@drecam.saclay.cea.fr},\mailto{luck@spht.saclay.cea.fr}}
\maketitle

\section{Introduction}

The non-equilibrium dynamics observed in a variety of systems,
ranging from glasses to granular media,
with its characteristic features of glassiness and aging~\cite{ange,glassy},
is often thought of as the motion of a particle in a complex energy landscape,
with many valleys and barriers~\cite{gold}.
This picture fully applies in the mean-field situation,
where barriers are infinitely high
and valleys appear as metastable states~\cite{tap,ks}.
For finite-dimensional systems with short-range interactions,
barrier heights and lifetimes are always finite at finite temperature,
so that metastability becomes a matter of time scales~\cite{gs,bir}.

The main goal of the present paper is to emphasize that the
zero-temperature dynamics of simple models such as Ising ferromagnets
provides, as an alternative to the mean-field situation,
interesting examples of dynamical systems with many attractors
(absorbing configurations, blocked configurations,
zero-temperature metastable states).
For completeness, we first give a brief review of the phenomenon
of metastability in the mean-field ferromagnet (Section~\ref{mft}),
and of the droplet picture implying that barrier heights become finite
for short-range interactions (Section~\ref{drop}).
We then turn to an overview of the configurational entropy
in disordered and complex systems,
and of the so-called Edwards hypothesis (Section~\ref{complex}).
The rest of this paper is devoted to zero-temperature single-spin-flip dynamics
of ferromagnetic Ising models.
We characterize the dynamics which lead to zero-temperature metastability,
and investigate the structure of the attractors thus obtained.
After a general presentation (Section~\ref{gene}),
we review the extensive amount of exact results available in one dimension
(\cite{ldt,prbr}, and especially~\cite{usrsa}) (Section~\ref{chain}),
and present original results on the square and honeycomb lattices
(Section~\ref{two}).
We end up with a brief discussion (Section~\ref{dis}).

\section{Metastability in the mean-field ferromagnet}
\label{mft}

The concepts of metastability and of spinodal line
date back to the early days of Thermodynamics,
with the seminal work of Gibbs~\cite{gibbs},
in the context of phase separation in simple systems such as fluids.
In order to summarize these early developments,
let us use the more modern language of the Landau theory
for a ferromagnet~\cite{ll}.
At the mean-field level,
i.e., neglecting any spatial dependence of the order parameter,
a ferromagnetic system of Ising type is
described by the Landau free energy density
\beq
F(M)=A(T-T_\c)\frac{M^2}{2}+B\frac{M^4}{4}-HM,
\label{gibbs}
\eeq
where $T$ is the temperature, $T_\c$ the Curie temperature,
$A$ and $B$ are positive phe\-no\-me\-no\-lo\-gi\-cal parameters,
and $H$ is the applied magnetic field.
The magnetization $M(T,H)$ is such that $F(M)$ is stationary.
It is therefore given by the equation of state
\beq
\frac{\dpar F}{\dpar M}=A(T-T_\c)M+BM^3-H=0.
\label{state}
\eeq
The corresponding magnetic susceptibility $\chi$ is such that
\beq
\frac{1}{\chi}=\frac{\dpar^2F}{\dpar M^2}=A(T-T_\c)+3BM^2.
\label{chi}
\eeq

At fixed temperature $T<T_\c$, the solutions to~(\ref{state})
are of the following three types:

\noindent (I) {\it stable}: absolute (global) minimum of the free energy,
for $\abs{M}>M_0(T)$,

\noindent (II) {\it metastable}: relative (local) minimum of the free energy,
for $M_\sp(T)<\abs{M}<M_0(T)$,

\noindent (III) {\it unstable}: maximum of the free energy,
for $\abs{M}<M_\sp(T)$.

\begin{figure}[htb]
\begin{center}
\includegraphics[angle=90,width=.5\linewidth]{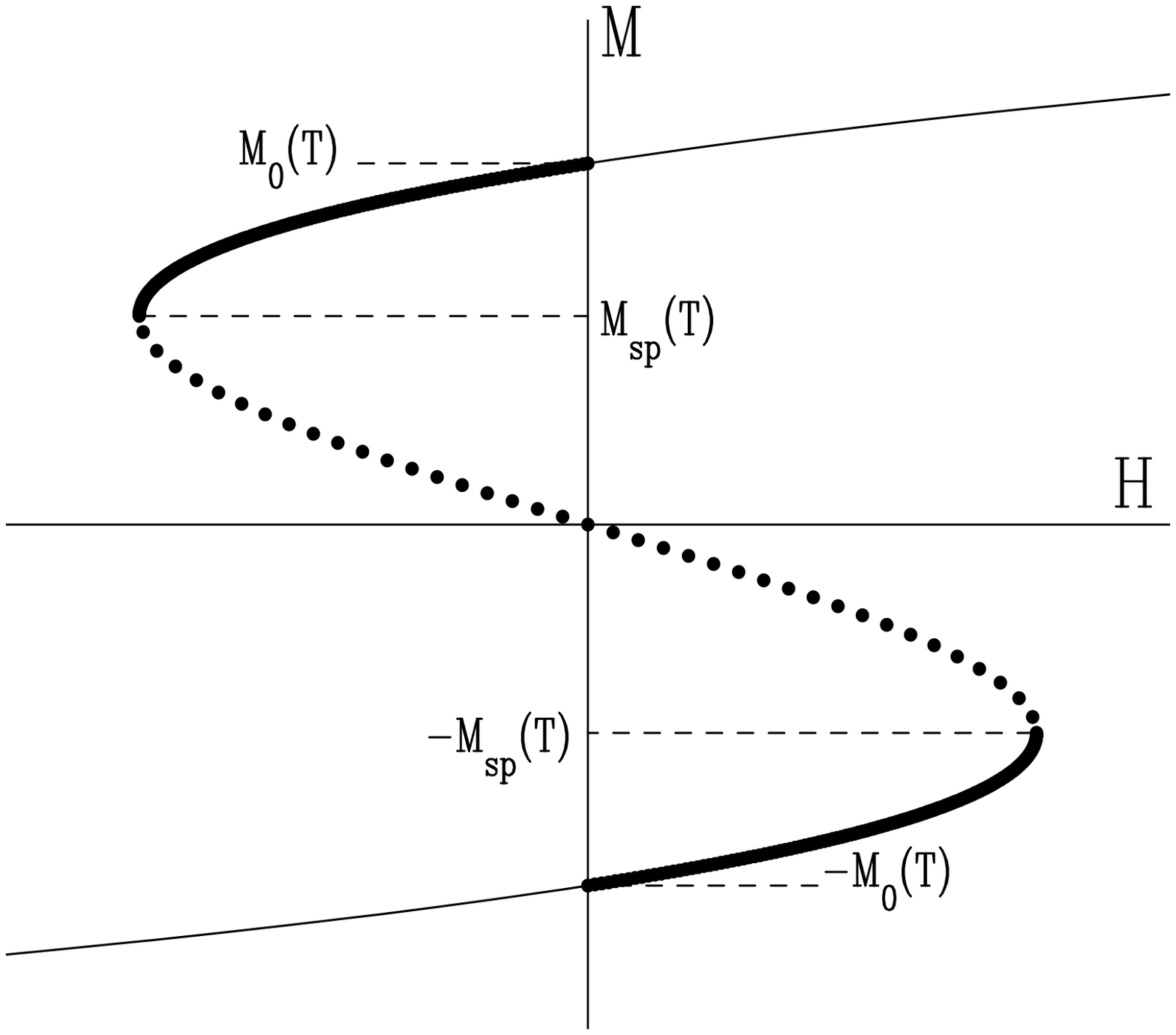}
\vskip 5mm
\includegraphics[angle=90,width=.5\linewidth]{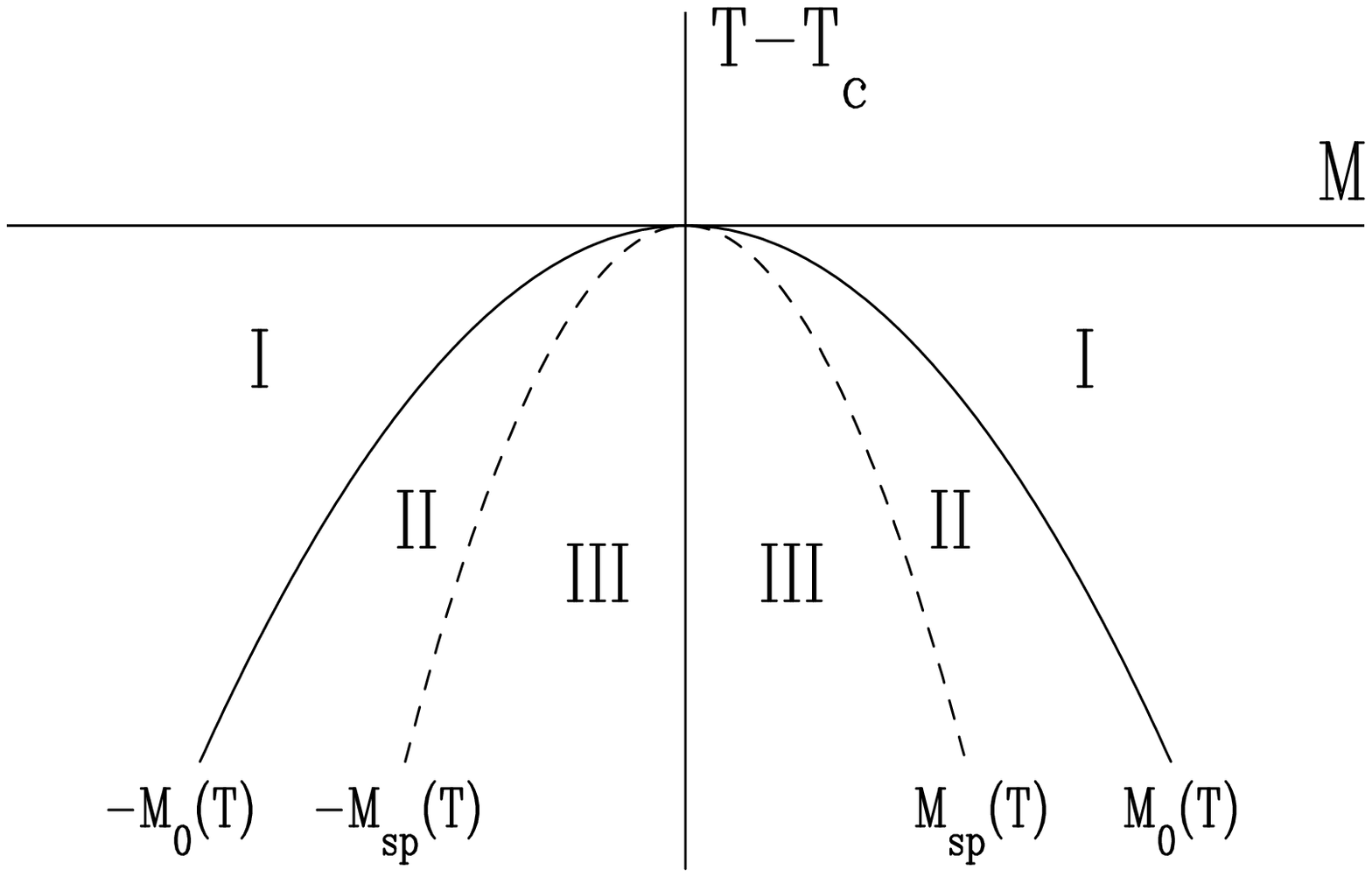}
\caption{\small
The three types of solutions to the equation of state~(\ref{state}).
Upper panel: Plot of the magnetization $M(H,T)$ against the magnetic field $H$,
for a fixed temperature $T<T_\c$.
Thin full line: stable solution (I).
Thick full line: metastable solution (II).
Thick dotted line: unstable solution (III).
Lower panel: Phase diagram in the $M$--$T$ plane.
Phases I to III correspond to the three types of solutions.
Full line: spontaneous magnetizations $M=\pm M_0(T)$.
Dashed line: spinodal magnetizations $M=\pm M_\sp(T)$.}
\label{figm}
\end{center}
\end{figure}

Figure~\ref{figm} illustrates this discussion.
When the magnetic field $H$ is positive,
the stable solution $M(H,T)$ describes the magnetization of the $+$ phase.
If $H$ decreases to zero,
the magnetization takes its spontaneous value $M(0,T)=M_0(T)$.
Coexistence is reached: the $+$ phase becomes degenerate with the $-$ phase.
If $H$ decreases further to
a small negative value, the $+$ phase still exists but becomes metastable.
Its magnetization $M(H,T)$ can be analytically continued
all the way until the endpoint of the metastability region,
which is reached for $\dpar F/\dpar M=\dpar^2F/\dpar M^2=0$.
The corresponding values of $M$ and $H$ are respectively called the spinodal
magnetization $M_\sp(T)$ and the coercive magnetic field $H_\co(T)$.
All physical quantities are singular at the spinodal endpoint.
The spontaneous magnetization, spinodal magnetization,
and coercive field read
\bea
M_0(T)=M(0,T)=\left(\frac{A(T_\c-T)}{B}\right)^{1/2},\nonumber\\
M_\sp(T)=M(-H_\co(T),T)
=\left(\frac{A(T_\c-T)}{3B}\right)^{1/2}=\frac{M_0(T)}{\sqrt{3}},\\
H_\co(T)=\left(\frac{4A^3(T_\c-T)^3}{27B}\right)^{1/2}
=\frac{2BM_0(T)^3}{3\sqrt{3}}.\nonumber
\eea

The metastable solution
is separated from the stable one by an extensive free energy barrier,
whose height $\Omega\,\Delta F$ grows proportionally to the volume
$\Omega$ of the system.
The activation energy density $\Delta F$ is the difference in $F(M,T)$
between the metastable solution (II) and the unstable one (III).
It has a finite limit
\beq
\Delta F=\frac{A^2(T_\c-T)^2}{4B}
\eeq
at coexistence, whereas it vanishes at the spinodal endpoint.

The continuation of the stable phase to the metastable one
corresponds to an analytical continuation in the mathematical sense.
The magnetization $M(H,T)$, defined as the stable solution to~(\ref{state})
for $H>0$, is indeed analytic at $H=0$.
The power-series expansion
\beq
M(H,T)=\sum_{n\ge0}a_n(T)\,H^n
\label{aims}
\eeq
has a finite radius of convergence, equal to $\abs{H_\co(T)}$.

\section{Finite-dimensional ferromagnet: Metastability vs.~droplet picture}
\label{drop}

It has been realized in the 1960s that metastability
is an artefact of the mean-field approximation.
For realistic Ising ferromagnetic systems with short-range interactions
in a finite-dimensional space,
Fisher~\cite{fisher} and Langer~\cite{langer}
have shown that the magnetization $M(H,T)$ of the $+$ phase,
or, equivalently, its free energy $F(H,T)$, is {\it not} analytic
at the coexistence point $H=0$.
Their argument relies on the droplet theory initiated by Frenkel~\cite{frenkel}.
Consider, in a $d$--dimensional ferromagnetic model below its Curie temperature,
a spherical droplet with radius $R$
of the $-$ phase ($M=-M_0(T)$) inside an infinite $+$ phase
($M=+M_0(T)$).
The corresponding excess free energy
\beq
f(R)=2\w R^dHM_0(T)+d\w R^{d-1}\s_0(T)
\eeq
is the sum of a volume term, proportional to the volume $\w R^d$ of the droplet,
to the spontaneous magnetization $M_0(T)$, and to the magnetic field $H$,
and of a surface term, proportional to the surface area $d\w R^{d-1}$
of the droplet and to the interfacial free energy $\s_0(T)$.
Here $\w=2\pi^{d/2}/\Gamma(d/2)$ denotes the volume of the
$d$--dimensional unit sphere.
If the magnetic field $H$ is positive,
the excess free energy is a positive and growing function of the radius $R$.
If the magnetic field is negative, however,
the excess free energy reaches a maximum~\cite{frenkel},
\beq
f_\c=f(R_\c)=\w\s_0(T)^d\left(\frac{d-1}{2\abs{H}M_0(T)}\right)^{d-1},
\eeq
for a finite critical radius
\beq
R_\c=\frac{(d-1)\s_0(T)}{2\abs{H}M_0(T)}.
\eeq

The existence of a critical droplet with a {\it finite} radius $R_\c$,
and hence of a {\it finite} barrier height $f_\c$,
has far-reaching consequences, both in statics and in dynamics.
The reduced barrier height $f_\c/T$ of the critical droplet
(in units where Boltzmann's constant is unity)
plays a role similar to that of the reduced instanton action
$S/\hbar$ in Quantum Mechanics~\cite{zinn}.

\begin{itemize}

\item {\it Statics.}
The free energy $F(H)$ of the $+$ phase is not analytic at $H=0$.
In other words, the formal power-series expansion~(\ref{aims}) is divergent:
its radius of convergence vanishes.
It can indeed be shown that $F(H)$ has a branch cut starting at $H=0$,
with an exponentially small imaginary part for $H<0$, of the form
\beq
\im F(H+\i0)\sim\exp\left(-\frac{f_\c}{T}\right).
\label{imf}
\eeq

\item {\it Dynamics.}
If the magnetic field is instantaneously turned from a positive
to a small negative value, the `metastable' $+$ phase
has a finite lifetime $\tau$,
whose order of magnitude is the inverse nucleation rate of a critical droplet,
given by an Arrhenius law in terms of the above barrier height:
\beq
\frac{1}{\tau}\sim\exp\left(-\frac{f_\c}{T}\right).
\label{tauinv}
\eeq

\end{itemize}

Near coexistence ($\abs{H}\to0$), the estimates~(\ref{imf})
and~(\ref{tauinv}) have an
exponentially small essential singularity of the form
\beq
\frac{1}{\tau}\sim\im F(H+\i0)\sim\exp\left(-\frac{\w\s_0(T)^d}{T}
\left(\frac{d-1}{2\abs{H}M_0(T)}\right)^{d-1}\right).
\eeq

To sum up, for ferromagnets and similar systems,
the metastability scenario {\it stricto sensu}
(analytic free energy at the coexistence point,
infinite lifetime in the metastable region, sharp spinodal line)
is a peculiar feature of the mean-field and of the zero-temperature situations.
For systems with short-range interactions,
at finite temperature in finite dimension,
the barrier height $f_\c$ is finite all over the would-be metastable region.
The free energy has an essential singularity at the coexistence point,
and the lifetime of the metastable phase is finite,
although it becomes expo\-nen\-tially large near coexistence.
The spinodal line therefore becomes a crossover between
the fast decay of an unstable phase (no barrier height)
and the slow decay of an approximately metastable phase (finite barrier height).
This crossover phenomenon has been recently investigated
by means of a sophisticated numerical approach~\cite{bustillos}.

\section{Metastable states and complexity in disordered and complex systems}
\label{complex}

From the 1970s on, a variety of complex systems,
such as structural glasses, spin glasses, and granular materials,
have been shown to possess many metastable states at low enough temperature,
or high enough density.
The number $\N(N;E)$ of metastable states with given energy density $E$
typically grows exponentially with the system size, as
\beq
\N(N;E)\sim\exp(N\,S_\conf(E)).
\label{nconf}
\eeq
The quantity $S_\conf(E)$ is referred to
as the configurational entropy, or complexity~\cite{sconf}.

Most investigations of the complexity
were motivated by dynamical considerations.
The dynamics of the above systems
in their low-temperature or high-density regime
often turns out to be so slow
that the system falls out of equilibrium~\cite{ange},
and exhibits aging and other characteristic features
of glassy dynamics~\cite{glassy}.
It has been proposed long ago to describe slow relaxational dynamics
in terms of the motion of a particle
in a complex energy (or free energy) landscape~\cite{gold},
with many valleys separated by barriers.
Several approaches have been developed,
in order to make this heuristic picture more precise,
and chiefly to identify these valleys.
Metastable states have thus been rediscovered under various names
and with various definitions, in different contexts:
valleys~\cite{ks}, pure states~\cite{tap,ktw},
inherent structures~\cite{sw}, quasi-states~\cite{fv}.

In the mean-field situation,
all these metastable states are separated by extensive barriers,
so that they have an infinite lifetime,
just as the unfavored phase in an Ising ferromagnet.
For finite-dimensional systems with short-range interactions,
barrier heights and lifetimes are always finite at finite temperature,
so that metastability becomes a matter of time scales.
Furthermore, from a dynamical viewpoint, the various concepts
of metastability recalled above are not equivalent~\cite{gs,bir}.

Whenever metastable states live forever,
i.e., either in the mean-field situation or at zero temperature,
a natural question concerns their {\it dynamical weights}.
Consider a system instantaneously quenched into its low-temperature
or high-density regime, from a randomly chosen initial configuration.
{\it Does the system sample all the possible metastable states
with given energy density with equal statistical weights,
i.e., with a uniform or flat measure, or, to the contrary,
does any detailed feature, such as the shape or size of the attraction basin
of each metastable state, matter?}
This question is of interest for any dynamical system having a large number
of attractors.

A similar question has been addressed in many works on the tapping
of granular systems~\cite{ldt,bm,pbt}.
Under tapping, a granular material constantly jumps
from a blocked configuration to a nearby one.
In this context, Edwards~\cite{edwards} proposed to describe situations
such as slow compaction dynamics,
or the steady-state dynamics under gentle tapping,
by means of a flat ensemble average over the {\it a priori ensemble}
of all the blocked configurations of the grains with prescribed density.

Extending the range of application of this idea far beyond its original scope,
the so-called {\it Edwards hypothesis} commonly refers to the assumption
that all the metastable states with given energy density are equivalent.
This hypothesis has two consequences.
First, the value of an observable can be obtained by a flat average
over the a priori ensemble, or Edwards ensemble, of all those metastable states.
Second, the configurational temperature, or Edwards temperature,
$T_\conf=(\d S_\conf/\d E)^{-1}$,
is expected to share the usual thermodynamical properties of a temperature,
and especially to coincide with the effective temperature
involved in the generalized fluctuation-dissipation formula
in the appropriate temporal regime~\cite{fdt}.
The concept of ergodicity, and the resulting thermodynamical construction,
therefore hold, as in equilibrium situations,
up to the replacement of configurations by metastable states,
and of temperature by the configurational temperature~$T_\conf$.

The Edwards hypothesis has been shown to be valid
for the slow relaxational dynamics of some mean-field models~\cite{fv},
where metastable states are indeed explored with a flat measure.
This hypothesis has also been found to hold for finite-dimensional systems,
at least as a good numerical approximation,
both in the steady-state of tapped systems
in the regime of weak tapping intensities~\cite{ldt,bm,pbt},
and in the regime of slow relaxational dynamics of various models~\cite{ba}.

Besides the mean-field geometry,
another physical situation where metastable states
are unambiguously defined is the zero-temperature limit,
where no barrier can be crossed at all.
The metastable states are then defined as
the blocked configurations under the chosen dynamics.
For instance, for an Ising model with single-spin-flip dynamics,
a metastable state is a configuration where each spin is aligned
with its local field, whenever the latter is non-zero.
The rest of this paper is devoted to the statistical analysis
of the blocked configurations thus generated,
before we come back to the so-called Edwards hypothesis in Section~\ref{dis}.

\section{Zero-temperature single-spin-flip dynamics of the Ising model:
Generalities}
\label{gene}

This section is devoted to an overview of zero-temperature
dynamics of ferromagnetic Ising models.
In the two following sections, we review the extensive available results
in one dimension (\cite{ldt,prbr}, and especially~\cite{usrsa}),
and present novel results in two dimensions.

The ferromagnetic Ising model is defined by the Hamiltonian
\beq
\H=-\sum_{(m,n)}\s_m\s_n,
\label{hamf}
\eeq
where the $\s_n=\pm1$ are classical Ising spins
sitting at the vertices of a regular lattice with coordination number~$z$,
and the sum runs over all pairs $(m,n)$ of nearest neighbors.

Consider single-spin-flip dynamics (i.e., Glauber dynamics, in a broad sense)
in continuous time.
Each spin is flipped ($\s_n\to-\s_n)$ with a rate $W(\del\H)$ per unit time.
This rate is assumed to only depend on the energy difference
implied in the flip,
\beq
\del\H=2h_n\s_n,
\eeq
where
\beq
h_n=-\frac{\dpar\H}{\dpar\s_n}=\sum_{m(n)}\s_m,
\eeq
is the local field acting on spin $n$.
The sum in the above definition runs over the $z$ nearest neighbors $m$
of site~$n$.

The condition of detailed balance at temperature $T=1/\beta$
with respect to the Hamiltonian $\H$ reads
\beq
\frac{W(\del\H)}{W(-\del\H)}=\exp(-\beta\del\H).
\label{deba}
\eeq
The two most usual choices of flipping rates,
or flipping probabilities in the case of discrete updates,
\beq
\matrix{
\hbox{Heat-bath:}\hfill& W_\HB(\del\H)=\frad{1}{1+\exp(2\beta\del\H)},
\hfill\cr
\hbox{Metropolis:}\quad\hfill& W_\M(\del\H)=\min(1,\exp(-\beta\del\H)),\hfill}
\label{hbm}
\eeq
obey the detailed balance condition~(\ref{deba}) at any temperature.
It is, however, less well-known that there is a vast family of dynamical rates,
besides these two choices, which obeys the condition~(\ref{deba}).
Indeed, the energy difference $\del\H$ may assume $z+1$
equidistant integer values:
\beq
\del\H=2z,\ 2z-4,\ \dots,\ -2z,
\label{list}
\eeq
so that there are a priori $z+1$ different rates.
If the coordination number $z$ is even (resp.~odd),
the condition~(\ref{deba}) only gives $z/2$ (resp.~$(z+1)/2$)
equations for the $z+1$ unknown rates,
so that there remain $(z+2)/2$ (resp.~$(z+1)/2$) free parameters
(one of them corresponding to the absolute time scale).

The arbitrariness described above persists in the zero-temperature limit.
The condition of detailed balance~(\ref{deba}) indeed reads
\beq
W(\del\H)=0\quad\hbox{for all }\del\H>0,
\label{baz}
\eeq
whereas nothing is imposed on the rates with $\del\H\le0$.
For the sake of definiteness,
we focus our attention onto the class of zero-temperature dynamics defined
by the following flipping rates:
\beq
W(\del\H)=\left\{\matrix{
0\hfill&\hbox{if }\del\H>0,\cr
W_0\quad\hfill&\hbox{if }\del\H=0,\cr
1\hfill&\hbox{if }\del\H<0.}\right.
\label{wz}
\eeq
The rate~$W_0$ relative to the {\it free} spins is kept as a parameter.
A free spin is defined as a spin $\s_n$
which is subjected to a zero local field ($h_n=0$, hence $\del\H=0$).
Free spins can only exist if the coordination number $z$ is even,
so that 0 belongs to the list~(\ref{list}).
The zero-temperature limits of the heat-bath and Metropolis rates~(\ref{hbm})
are respectively ${W_0}_\HB=1/2$ and ${W_0}_\M=1$.

The zero-temperature dynamics of ferromagnetic Ising systems
is far from being a trivial problem in general~\cite{red,ns}.
The single-spin-flip dynamics defined by~(\ref{wz}) is a descent dynamics
(i.e., every move strictly lowers the total energy)
in the following two situations:

\begin{itemize}

\item $z$ is odd, so that there are no free spins at all.
\item $z$ is even and $W_0=0$.
The corresponding dynamics is said to be {\it constrained}.
As such it belongs to the class of
{\it kinetically constrained} systems~\cite{barc,riso}.

\end{itemize}

In both situations, every spin flips a finite number of times,
and the system gets trapped in a non-trivial attractor
(absorbing configuration, blocked confi\-guration,
zero-temperature metastable state).

\section{Constrained zero-temperature single-spin-flip dynamics
on the Ising chain}
\label{chain}

The one-dimensional situation of a chain of spins corresponds to the
smallest possible value of the coordination number ($z=2$),
so that the list of values of the energy difference is 4, 0, and $-4$.
The heat-bath dynamics has been solved exactly by Glauber~\cite{glau}.

Let us focus our attention onto zero-temperature dynamics.
For any non-zero value of the rate $W_0$ corresponding to free spins,
the dynamics belongs to the universality class of the zero-temperature
Glauber model.
This is a prototypical example of phase ordering by domain growth
(coarsening)~\cite{bray}.
The typical size of ordered domains of consecutive $+$ and $-$ spins
grows as $L(t)\sim t^{1/2}$.
The energy density $E(t)=-1+2/L(t)$ therefore
relaxes to its ground-state value as $t^{-1/2}$.
The particular value $W_0=0$ corresponds to
the constrained zero-temperature Glauber dynamics,
which has been investigated in~\cite{ldt,prbr}, and especially in~\cite{usrsa}.
Hereafter we summarize the main results of these investigations.

\subsection{Mapping onto the dimer RSA model}

In the constrained zero-temperature Glauber dynamics,
the only possible moves are flips of isolated spins:
\beq
-+-\;\to\;---,\quad+-+\;\to\;+++.
\label{gls}
\eeq
Each move suppresses two consecutive unsatisfied bonds.
The system therefore eventually reaches a blocked configuration,
where there is no isolated spin, i.e., up and down spins form clusters
whose length is at least two.
Equivalently, each unsatisfied bond (or domain wall) is isolated.
These blocked configurations are the zero-temperature analogues
of metastable states.

The dynamics can be recast in the following illuminating way.
Going to the dual lattice, where dual sites represent bonds,
let us represent unsatisfied bonds as empty dual sites,
and satisfied bonds as occupied dual sites:
\beq
\left\{\matrix{
\t_n=\s_n\s_{n+1}=-1\lra\bl,\hfill\cr
\t_n=\s_n\s_{n+1}=+1\lra\no,\hfill
}\right.
\label{corrbond}
\eeq
so that the moves~(\ref{gls}) read
\beq
\bl\bl\to\no\no.
\label{rsadim}
\eeq
This mapping shows at once that the dynamics is fully irreversible,
in the sense that each spin flips at most once
during the whole history of the sample.

The dynamics~(\ref{rsadim}) is identical to that of the
random sequential adsorption (RSA) of lattice dimers,
which has been considered long ago~\cite{flory,core}.
This is one of the simplest examples of RSA problems,
for which powerful analytical techniques are available in one dimension,
both on the lattice and in the continuum~\cite{rsa}.
A blocked configuration thus appears as a jammed state of the dimer RSA model,
as illustrated on the following example:
\bea
&&+---++--++++---+++-----+++--\cr
&&{\hskip 2.5mm}
\bl\no\no\bl\no\bl\no\bl\no\no\no\bl\no\no\bl
\no\no\bl\no\no\no\no\bl\no\no\bl\no
\eea

\subsection{Relaxation of the mean energy}

We consider as initial state the equilibrium state at finite temperature $T_0$.
Each bond variable is independently drawn from the binary distribution
\beq
\t_n=\left\{\matrix{
-1\hfill&(\bl)\hfill&\hbox{with probability}\;p,\hfill\cr
+1\hfill&(\no)\hfill&\hbox{with probability}\;1-p,\hfill
}\right.
\label{epsf}
\eeq
where the parameter $p$ is related to the energy density $E_0$
of the initial state and to the corresponding temperature $T_0$ by
\beq
\beta_0=\frac{1}{T_0}=\frac{1}{2}\,\ln\frac{1-p}{p},\quad E_0=-1+2p.
\eeq

It is a common feature of one-dimensional RSA problems~\cite{rsa}
that the densities of certain patterns obey closed rate equations.
In the present case,
the densities $p_\ell(t)$ of clusters made of exactly~$\ell$
consecutive unsatisfied bonds (empty dual sites)
obey the linear equations~\cite{usrsa,flory}
\beq
\frac{\d p_\ell(t)}{\d t}=-(\ell-1)p_\ell(t)+2\sum_{k\ge\ell+2}p_k(t)
\label{dpl}
\eeq
for $\ell\ge1$, with $p_\ell(0)=(1-p)^2p^\ell$.
These equations can be solved by making the Ansatz
$p_\ell(t)=a(t)\,z(t)^\ell$ for $\ell\ge1$.
The solution thus obtained,
\beq
p_\ell(t)=(1-p\e^{-t})^2\exp(2p(\e^{-t}-1))\,p^\ell\e^{-(\ell-1)t},
\label{plt}
\eeq
leads to
\beq
E(t)=-1+2\sum_{\ell\ge1}\ell\,p_\ell(t)=-1+2p\exp(2p(\e^{-t}-1)).
\eeq
Only clusters of length $\ell=1$ survive in the blocked configurations,
as could be expected.
These defects occur with a finite density,
\beq
p_1(\infty)=p\e^{-2p},
\eeq
so that the energy density of blocked configurations
takes a non-trivial value, which depends continuously
on the initial temperature via the parameter $p$~\cite{ldt,prbr,usrsa}:
\beq
E_\infty=E(\infty)=-1+2p\e^{-2p}.
\label{einf}
\eeq
This non-trivial dependence is a clear evidence
that the dynamics is not ergodic.
For an initial state close to the ferromagnetic ground-state
($E_0\to-1$, i.e., $p\to0$), the behavior $E_\infty\approx E_0-4p^2$
is easily explained in terms of the clusters made of two empty sites.
The energy of blocked states then increases monotonically with $p$,
up to the maximum value $E_\infty=-1+\e^{-1}\approx-0.632121$,
for an uncorrelated (infinite-temperature)
initial state ($p=1/2$, i.e., $E_0=0$),
and then decreases monotonically with $p$, down to the value
$E_\infty=-1+2\e^{-2}\approx-0.729329$,
corresponding to an antiferromagnetically ordered initial state ($p=1$).

\subsection{Distribution of the blocking time}

The late stages of the dynamics are dominated
by an exponentially small density of surviving clusters made of two empty sites.
More precisely,~(\ref{plt}) shows that their density reads
$p_2(t)\approx\alpha\e^{-t}$, with $\alpha=p^2\e^{-2p}$.
The dynamics can therefore be effectively described by
a collection of $\alpha N$ such clusters,
each cluster decaying exponentially with unit rate, when a down spin flips.
The blocking time $T_N$ is the largest of the decay times of those clusters.
For a large sample, it is therefore distributed according to extreme-value
statistics~\cite{gumbel}:
\beq
T_N=\ln(\alpha N)+X_N,
\label{tx}
\eeq
where the fluctuation $X_N$ remains of order unity,
and is asymptotically distributed according to the Gumbel law:
$f(X)=\exp(-X-\e^{-X})$.

\subsection{Distribution of the final energy: Dynamics vs.~a priori ensemble}

The attractors of the zero-temperature constrained
single-spin-flip dynamics of the Ising chain are the spin configurations
where every unsatisfied bond is isolated.
The most natural statistical description of these attractors
is provided by the a priori ensemble
where all the blocked configurations with given energy density $E$
are taken with equal weights.

In this section, the exact distribution of the final energy
of the blocked con\-fi\-gu\-ra\-tions is compared to the prediction
of the a priori approach.
It turns out that there are two a priori predictions,
the second (refined) one being a far better approximation
than the first (naive) one.

For a finite chain made of $N$ spins,
the number of blocked configurations with exactly $M$ unsatisfied bonds,
i.e., empty dual sites, reads
\beq
\N_{N,M}=\bin{N-M+b}{M},
\label{bino}
\eeq
where $b=0$ (resp.~1) for a closed (resp.~open) chain.
This is indeed the number of ways of inserting the $M$ empty sites
in the $N-M+b$ spaces made available by the presence
of $N-M$ occupied sites, with at most one empty site
per available space~\cite{usrsa}.
In the limit of a large system ($M,N\to\infty$,
with a fixed ratio $M/N=(E+1)/2$), this number grows exponentially,
irrespective of $b$, as
\beq
\N(N;E)\sim\exp(N\,S_\ap(E)),
\eeq
in accord with~(\ref{nconf}).
The a priori configurational entropy reads~\cite{ldt,cri,bfs}
\beq
S_\ap(E)=E\ln(-2E)+\frac{1-E}{2}\ln(1-E)-\frac{1+E}{2}\ln(1+E).
\label{sap}
\eeq

This result can be alternatively derived
by means of the transfer-matrix method~\cite{baxter,tm}.
For a finite chain of $N$ spins, we introduce the partition function
\beq
Z_N=\sum_\C\e^{-\g\H(\C)},
\eeq
where the sum runs over all the blocked configurations $\C$,
as well as the partial partition functions $Z_N^\no$, $Z_N^\bl$
labeled by the prescribed value of the last bond, according to~(\ref{corrbond}).
The latter quantities obey the recursion
\beq
\pmatrix{Z_{N+1}^\no\cr Z_{N+1}^\bl}=\T\pmatrix{Z_N^\no\cr Z_N^\bl},\quad
\T=\pmatrix{\e^\g&\e^\g\cr\e^{-\g}&0}.
\eeq
The $2\times2$ transfer matrix $\T$ has eigenvalues
\beq
\lam_\pm=\frac{\e^\g\pm\sqrt{4+\e^{2\g}}}{2},
\eeq
so that $Z_N^\no\sim Z_N^\bl\sim\exp(N\ln\lam_+)$.
The configurational entropy $S_\ap(E)$
is therefore given by a Legendre transform:
\beq
\ln\lam_+(\g)+\g E-S_\ap(E)=0,
\quad E=-\frac{\d\ln\lam_+}{\d\g},
\quad\g=\frac{\d S_\ap}{\d E},
\label{thertrans}
\eeq
hence
\beq
E=-\frac{\e^\g}{\sqrt{4+\e^{2\g}}},\quad\g=\ln\frac{-2E}{\sqrt{1-E^2}},
\label{egge}
\eeq
so that~(\ref{sap}) is recovered.

The most naive prediction for the probability distribution $f(E)$
of the final energy density
is that $f(E)$ is proportional to the number $\N(N;E)$.
We thus obtain a large-deviation estimate of the form
\beq
f(E)\sim\exp(-N\,\S_\ap(E)),
\eeq
where the naive prediction for the large-deviation function $\S_\ap(E)$ reads
\beq
\S_\ap(E)=S_\max-S_\ap(E),
\label{signai}
\eeq
with
\beq
S_\max=\ln\frac{\sqrt{5}+1}{2}\approx0.481212
\label{smax}
\eeq
being the maximum of the configurational entropy $S_\ap(E)$.
This maximum is reached for
\beq
E_\max=-\frac{1}{\sqrt{5}}\approx-0.447214,
\label{eap}
\eeq
which is therefore the typical a priori energy density
of a blocked configuration.

A more refined prediction for the distribution of the final energy density,
introduced in the context of Kawasaki dynamics~\cite{uskawa},
resides on the following observation.
The mean energy density of blocked configurations,
as they are generated by the zero-temperature dynamics,
is equal to $E_\infty$, given by~(\ref{einf}),
which is in general different from the a priori value~(\ref{eap}).
Within the a priori formalism under consideration,
this difference is taken into account
by attributing to every configuration $\C$
an extra weight of the form $\exp(-\g_{\rm eff}NE(\C))$.
The effective inverse temperature $\g_{\rm eff}$ is chosen so that
the mean energy density
coincides with~(\ref{einf}), hence $\g_{\rm eff}=\g(E_\infty)$,
where $\g(E)=\d S_\ap/\d E$ is given in~(\ref{egge}).
This procedure amounts to replacing the configurational entropy $S_\ap(E)$,
which is maximal at $E=E_\max$, by the relative entropy
\beq
S_\ap(E\vert E_\infty)=S_\ap(E)-(E-E_\infty)\g(E_\infty),
\eeq
which is maximal at $E=E_\infty$ by construction.
The resulting prediction for the distribution
of the final energy density now involves
the refined large-deviation function
\beq
\S_\ap(E\vert E_\infty)=S_\ap(E_\infty)-S_\ap(E)+(E-E_\infty)\g(E_\infty).
\label{sigref}
\eeq

It turns out that the distribution of the final energy density
of the blocked configurations,
as they are generated by the zero-temperature dynamics,
has been evaluated exactly by analytical means~\cite{usrsa}.
We prefer to skip every detail of this rather lengthy derivation
and to state the result.
The distribution of the final energy density
is given by an exponential estimate of the form
\beq
f(E)\sim\exp(-N\,\S_\dyn(E)),
\eeq
where the large-deviation function $\S_\dyn(E)$
depends on the parameter $p$ characterizing the initial state,
and reads, in parametric form:
\bea
&&{\hskip -4.75mm}
\S_\dyn=\ln z+\frac{(1+(2p-1)z)^2-(z-1)^2\e^{4p z}}
{4p z^2\e^{2p z}(2(1-p)+(2p-1)z)}
\ln\frac{1+(2p-1)z+(1-z)\e^{2p z}}{1+(2p-1)z-(1-z)\e^{2p z}},\nonumber\\
&&
E=-1+\frac{(1+(2p-1)z)^2-(z-1)^2\e^{4p z}}
{2p z^2\e^{2p z}(2(1-p)+(2p-1)z)}.
\label{sigf}
\eea

\begin{figure}[htb]
\begin{center}
\includegraphics[angle=90,width=.5\linewidth]{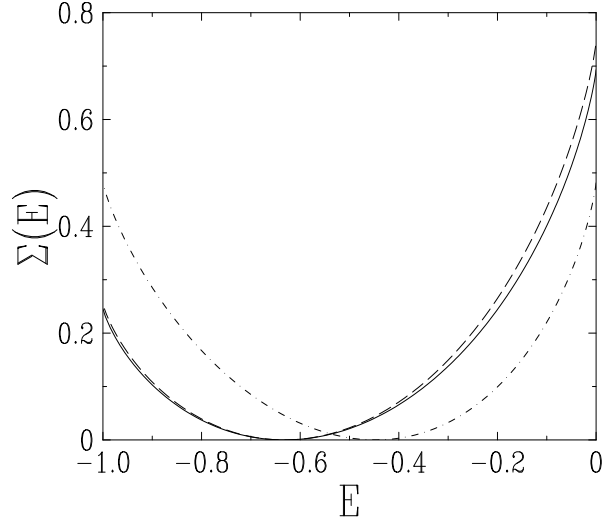}
\caption{\small
Full line: plot of the dynamical entropy
of the ferromagnetic chain with constrained Glauber dynamics,
given by $\S_\dyn(E)$~(\ref{sigf}), against energy $E$,
for an uncorrelated initial state ($p=1/2)$.
Dash-dotted line: prediction~(\ref{signai}) of the naive a priori approach.
Dashed line: prediction~(\ref{sigref}) of the refined a priori approach
(after~\cite{usrsa}).}
\label{figent}
\end{center}
\end{figure}

Figure~\ref{figent} shows a comparison between the dynamical entropy
$\S_\dyn(E)$, given by~(\ref{sigf}),
for an uncorrelated (infinite-temperature) initial state ($p=1/2)$,
and the predictions~(\ref{signai}) of the naive a priori approach
and~(\ref{sigref}) of refined a priori approach.
The refined prediction turns out to be a good approximation
to the dynamical entropy, although it is not exact.
The following particular values allow a quantitative comparison.
At the ground-state energy density $E=-1$, one has
\beq
\S_\dyn(-1)=\ln z_\c(p),\quad
\S_\ap(-1\vert E_\infty)=\ln\frac{1-p\e^{-2p}}{1-2p\e^{-2p}}
\eeq
with
\beq
1+(2p-1)z_\c+(1-z_\c)\e^{2p z_\c}=0,
\label{zcp}
\eeq
i.e., for $p=1/2$,
$\S_\dyn(-1)\approx0.245660$, $\S_\ap(-1\vert E_\infty)\approx0.255408$.
At the maximum allowed energy $E=0$, one has
\beq
\S_\dyn(0)=-\frac{1}{2}\ln p(1-p),\quad
\S_\ap(0\vert E_\infty)=\frac{1}{2}\ln\frac{1-p\e^{-2p}}{p\e^{-2p}},
\eeq
i.e., for $p=1/2$,
$\S_\dyn(0)\approx0.693147$, $\S_\ap(0\vert E_\infty)\approx0.744940$.

Another quantity of interest is the dynamical specific heat $C_\dyn$,
which is defined as the curvature of the dynamical entropy
at the mean energy density,
\beq
\S_\dyn(E)\approx\frac{(E-E_\infty)^2}{2C_\dyn},
\eeq
so that the energy variance (in the sense of sample to sample fluctuations
of the observed energy per spin in the blocked configurations) behaves as
\beq
\var E\approx\frac{C_\dyn}{N}.
\eeq
The exact dynamical entropy~(\ref{sigf})
and the refined a priori prediction~(\ref{sigref}) respectively yield
\beq
C_\dyn=4p\e^{-4p}(1-p+4p^2),\quad
C_\ap=4p\e^{-2p}(1-p\e^{-2p})(1-2p\e^{-2p}),
\eeq
i.e., for $p=1/2$,
$C_\dyn\approx0.406006$, $C_\ap\approx0.379540$.

\subsection{Correlations: Dynamics vs.~a priori ensemble}

Let us define the spin correlation function $C_n$
and the connected energy correlation function $G_n$ as
\beq
C_n=\dmean{\s_0\s_n},\quad
G_n=\dmean{\t_0\t_n}-E^2=\dmean{\s_0\s_1\s_n\s_{n+1}}-E^2.
\eeq

This section is devoted to a comparison between the exact
energy correlation function $G_n$
and the prediction of the refined a priori description.
It turns out that the spin correlation function $C_n$
is more difficult to handle,
both from the a priori and from the dynamical viewpoint.

The energy correlation function $(G_n)_\ap$
in the a priori ensemble at fixed energy density $E$
can again be evaluated by the transfer-matrix method.
We have, for~$n\ge0$ in the bulk of an infinitely long chain,
\beq
(G_n)_\ap
=\langle L_+\vert\En\vert R_-\rangle\langle L_-\vert\En\vert R_+\rangle
\left(\frac{\lam_-}{\lam_+}\right)^n.
\eeq
In this expression, $\En=\diag(-1,+1)$ is the energy operator, while
\beq
\langle L_\pm\vert=\frac{1}{\lam_\pm^2+1}\pmatrix{\lam_\pm&\e^\g},\quad
\vert R_\pm\rangle=\pmatrix{\e^\g\lam_\pm\cr1}
\eeq
are the left and right eigenvectors of $\T$
associated with the eigenvalues $\lam_\pm$.
We have consistently $\langle L_+\vert\En\vert R_+\rangle=E$.
After some algebra we obtain the following expression,
involving only the mean energy~$E$~\cite{ldt}:
\beq
(G_n)_\ap=(1-E^2)\left(-\frac{1+E}{1-E}\right)^n.
\label{cr}
\eeq
The connected energy correlation function in the a priori ensemble
thus exhibits an exponential fall-off, modulated by an oscillating sign.

The energy (i.e., occupancy, in the RSA language) correlation function
of the blocked configurations,
as they are generated by the zero-temperature dynamics,
has also been evaluated exactly~\cite{usrsa}.
Skipping again every detail, one has
\beq
(G_n)_\dyn
=2p\e^{-2p}\left((1-2p)\frac{(-2p)^n}{n!}
-2p\sum_{m\ge n+1}\frac{(-2p)^m}{m!}\right).
\label{cfinf}
\eeq
Whenever $p\ne1/2$, i.e., $T_0\ne\infty$,
the first term is leading, hence $(G_n)_\dyn\sim(-2p)^n/n!$.
In the case of an infinite initial temperature ($p=1/2$),
one has $(G_n)_\dyn\sim(-1)^n/(n+1)!$.
The connected energy correlations therefore exhibit a factorial decay,
modulated by an oscillating sign.

\section{Zero-temperature single-spin-flip dynamics in two dimensions:
Honeycomb vs.~square lattice}
\label{two}

Leaving aside the one-dimensional realm,
where exact analytical results are available,
by means of an exact mapping of the dynamics onto an RSA problem,
let us now turn to some novel results pertaining to two-dimensional examples
of the zero-temperature single-spin-flip dynamics
described in Section~\ref{gene}.
We will investigate the following cases:

\begin{itemize}

\item
{\it Honeycomb lattice}.
The coordination number $z=3$ is odd,
so that the standard zero-temperature dynamics~(\ref{wz}) is a descent dynamics.

\item 
{\it Square lattice.}
The coordination number $z=4$ is even,
so that the situation is qualitatively similar to that of the chain.
The dynamics~(\ref{wz}) is a bona fide coarsening dynamics
for any non-zero value of the rate $W_0$,
whereas the constrained dynamics ($W_0=0$) is a descent dynamics.

\end{itemize}

\begin{figure}[htb]
\begin{center}
\includegraphics[angle=90,width=.48\linewidth]{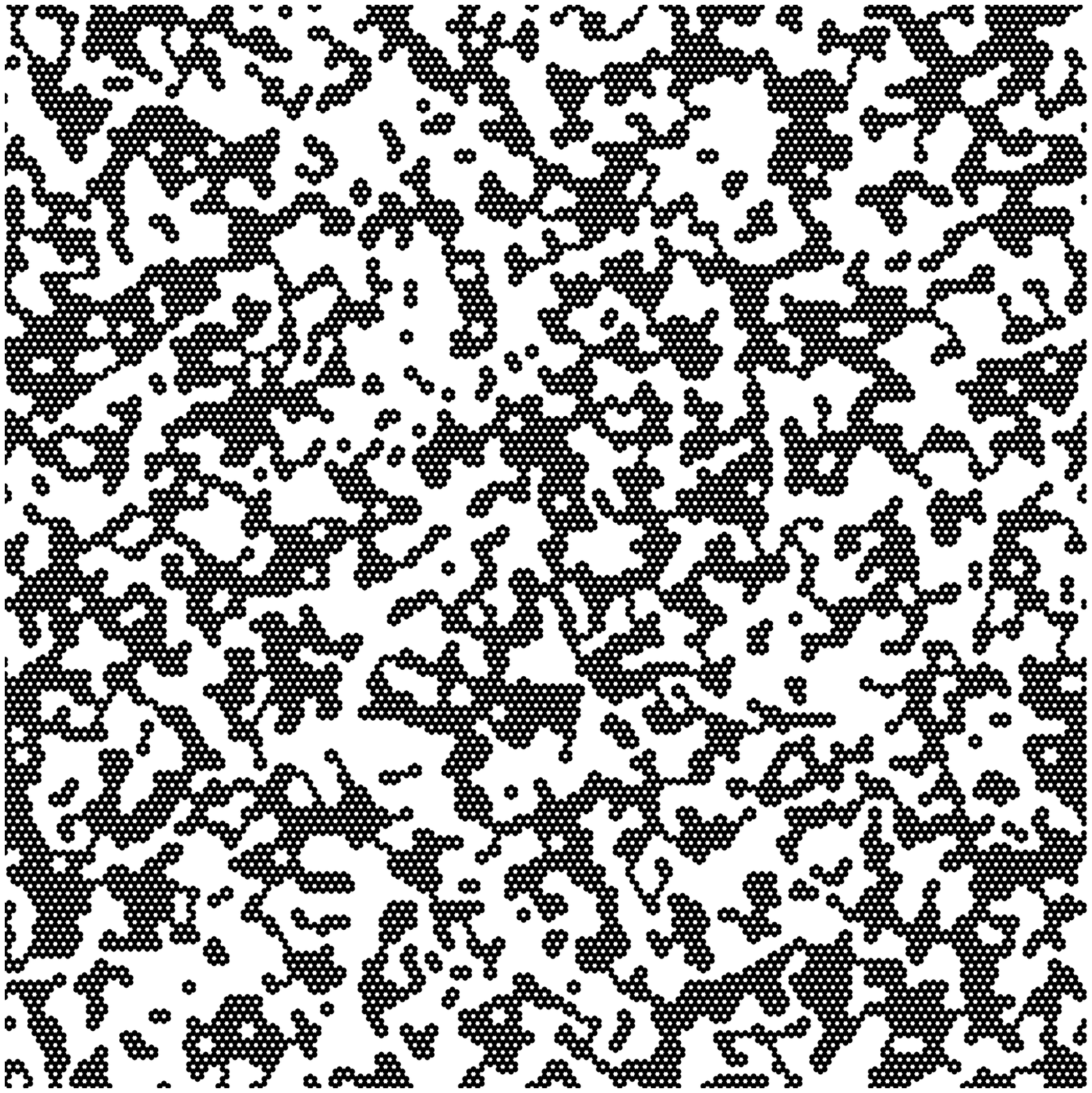}
{\hskip 1mm}
\includegraphics[angle=90,width=.48\linewidth]{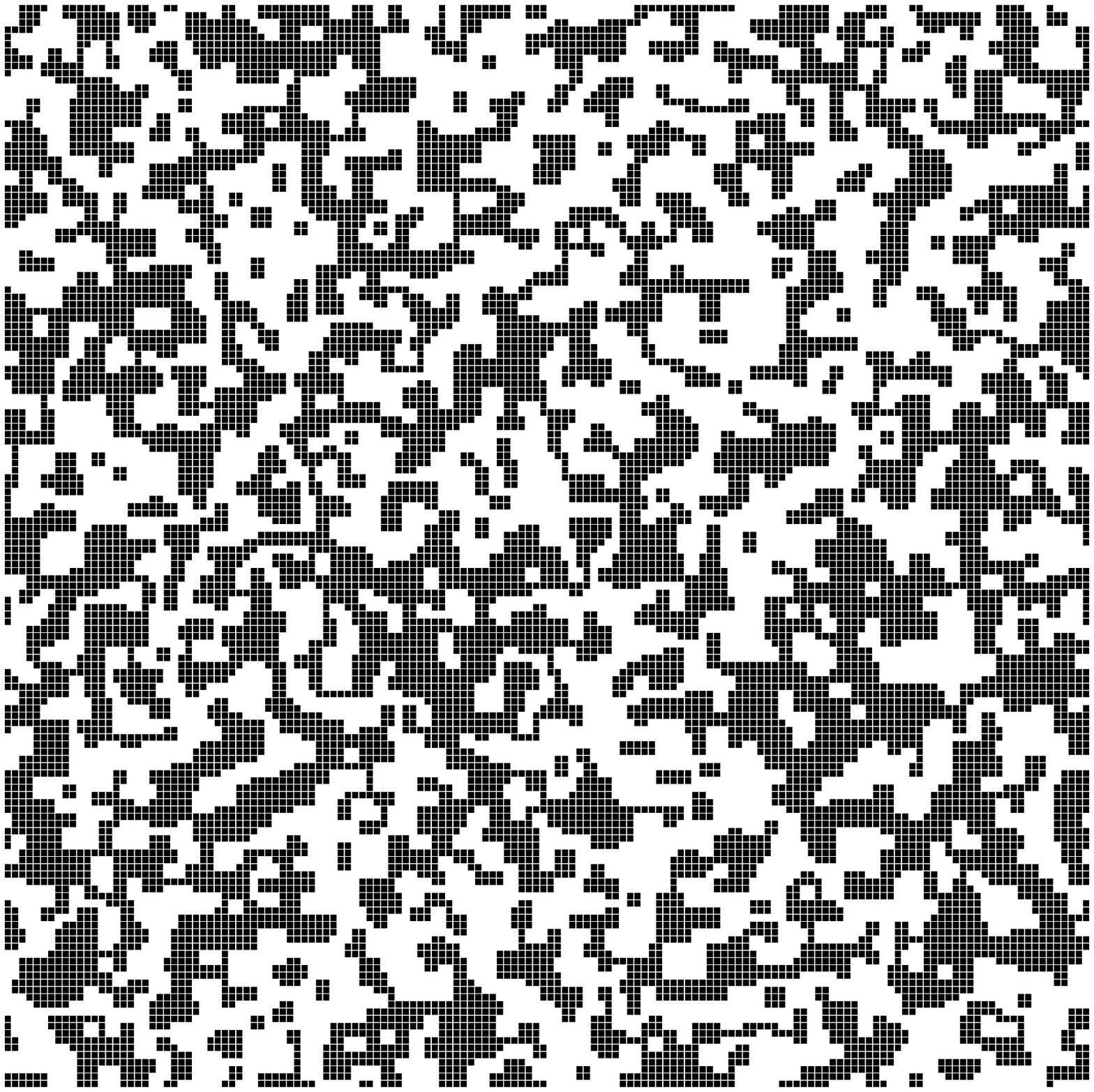}
\caption{\small
Typical blocked configuration generated by the zero-temperature
single-spin-flip dynamics considered here.
Black (resp.~white) sites represent up (resp.~down) spins.
Left: dynamics on the honeycomb lattice ($z=3$ is odd).
Right: constrained dynamics on the square lattice ($z=4$ is even, $W_0=0$).}
\label{fighc}
\end{center}
\end{figure}

In the following we discuss in parallel
the standard zero-temperature dynamics on the honeycomb lattice
and the constrained one on the square lattice.
Both dynamics lead to metastability:
the system gets trapped in a finite time into one
of many blocked configurations (zero-temperature metastable states).
Figure~\ref{fighc} shows typical blocked configuration
generated by both dynamics described above on samples of $150^2$ spins.

\subsection{Distribution of the number of flips}

The constrained single-spin-flip dynamics on the Ising chain
described in Section~\ref{chain} is fully irreversible,
in the sense that every spin flips at most once
during the whole history of the system,
before a blocked configuration is reached.

\begin{figure}[htb]
\begin{center}
\includegraphics[angle=90,width=.5\linewidth]{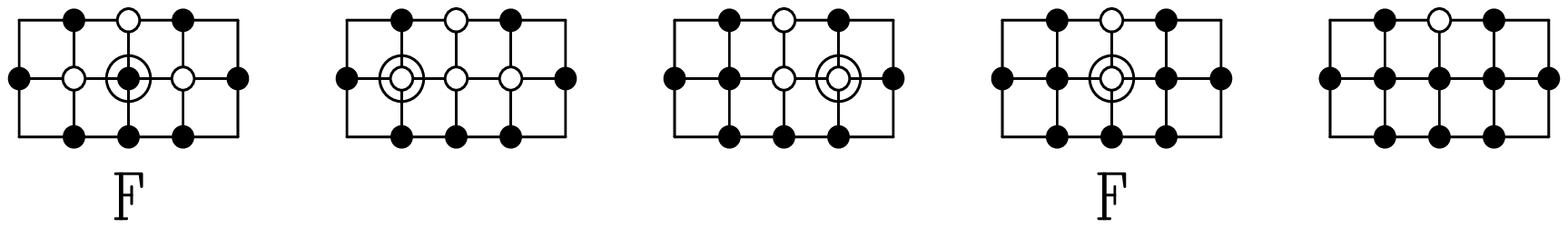}
\vskip 8mm
\includegraphics[angle=90,width=.5\linewidth]{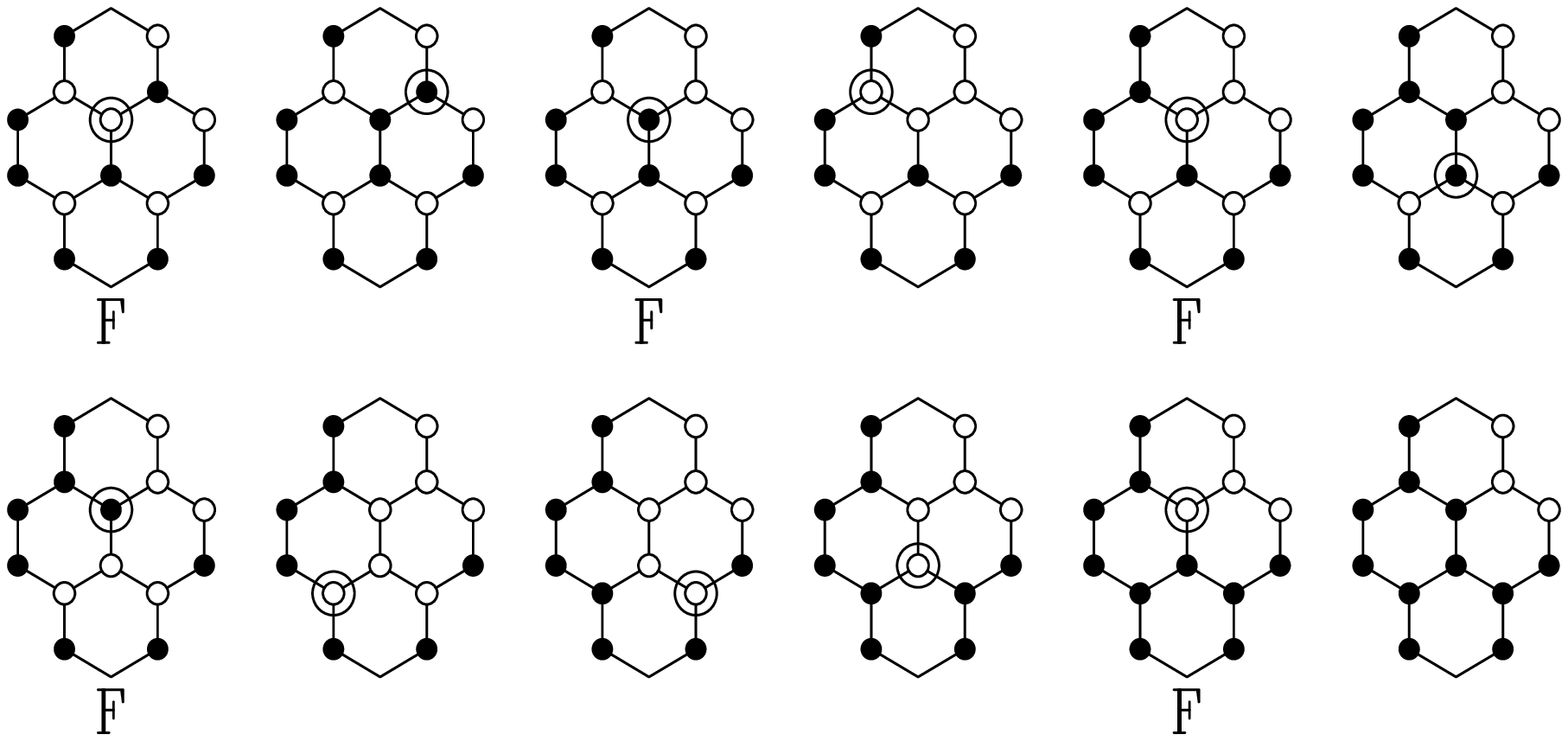}
\caption{\small
Examples of histories such that a given spin $\s_0$
experiences the maximum allowed number of flips.
At each step the spin about to flip is circled.
The letter F marks those steps where $\s_0$ is about to flip.
Upper panel (left to right): square lattice:
$\s_0$ (at center) flips $M_\ca=2$ times.
Lower panel (top to bottom and left to right): honeycomb lattice:
$\s_0$ (at bottom of uppermost hexagon) flips $M_\he=5$ times.}
\label{figflip}
\end{center}
\end{figure}

In the present situation,
although the two dynamics considered are descent dynamics in a global sense
(the total energy decreases at each spin flip),
a given spin can flip more than once.
It turns out that the maximum number of flips is $M_\ca=2$
for the constrained dynamics on the square lattice,
and $M_\he=5$ for the dynamics on the honeycomb lattice.
Skipping details,
let us mention that these bounds can be derived by considering the number
of unsatisfied bonds around the four squares,
or the three hexagons, which surround a given spin.
Figure~\ref{figflip} shows an example of a spin whose history
contains this maximum number of flips in each case.
The mean number of flips of a given spin remains however well below unity,
at least for uncorrelated initial configurations.
For dynamics on the honeycomb lattice,
the mean number of flips is $\dmean{M}_\he\approx0.4244$,
while the fraction of spins which flip five times
is of order $p_\he(5)\sim10^{-8}$.
For constrained dynamics on the square lattice,
the mean number of flips is $\dmean{M}_\ca\approx0.2577$,
while the fraction of spins which flip twice is $p_\ca(2)\approx0.00203$.

\subsection{Distribution of the blocking time}

\begin{figure}[htb]
\begin{center}
\includegraphics[angle=90,width=.5\linewidth]{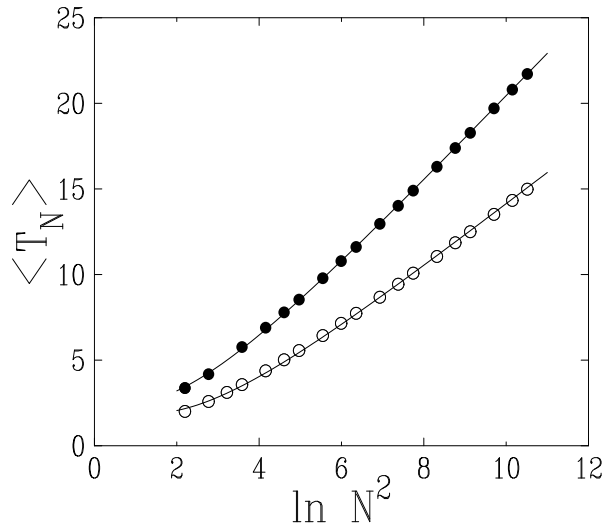}
\caption{\small
Plot of the mean blocking time $\dmean{T_N}$
of zero-temperature single-spin-flip dynamics,
against the number of spins $N^2$.
Full symbols: standard dynamics on the honeycomb lattice.
Empty symbols: constrained dynamics on the square lattice.
Full lines: fits (see text) with asymptotic slopes
$\t_\he\approx2.49$ and $\t_\ca\approx1.84$.}
\label{figtemps}
\end{center}
\end{figure}

We have measured by means of numerical simulations
the blocking time of finite samples of size $N\times N$
for the descent dynamics on the honeycomb and square lattice described above,
with uncorrelated random initial configurations.
Figure~\ref{figtemps}
shows a plot of the mean blocking time $\dmean{T_N}$ against the size
(number of spins $N^2$) of the simulated samples.
The data are well fitted by the form $\dmean{T_N}=a\ln N^2+b+c/N$.

This observation suggests that the late stages of the descent dynamics
can again be described in an effective way
by a dilute population of two-level systems,
which relax independently of each other,
with a common characteristic time $\t$,
so that their density falls off exponentially, as $p(t)\sim\exp(-t/\t)$.
Extreme-value statistics indeed implies
\beq
\dmean{T_N}\approx\t\ln N^2.
\label{evs}
\eeq
The fits shown in Figure~\ref{figtemps} yield
the characteristic times $\t_\he\approx2.49$ for the honeycomb lattice,
and $\t_\ca\approx1.84$ for the square lattice.
The fact that these estimates do not identify with simple numbers,
together with the rather large observed correction to the result~(\ref{evs}),
fitted by a $1/N$ term,
suggests however that there might be more than one type of two-level systems.

\subsection{Spin correlations}

We have measured the on-axis spin correlation function
$C_n=\dmean{\s_\vecz\s_\vecn}$,
where $\vecn$ denotes the site at distance $n$ lattice bonds
from the origin $\vecz$ along one of the main axes
of either lattice, and where $\dmean{\dots}$
again denotes an average over the blocked configurations.
The spin correlation at unit distance $C_1$
is nothing but minus the mean energy density
$E_\infty$ of the blocked configurations per bond.
We obtain $(C_1)_\he=-(E_\infty)_\he\approx0.7485$ for the honeycomb lattice,
and $(C_1)_\ca=-(E_\infty)_\ca\approx0.6126$ for the square lattice.

\begin{figure}[htb]
\begin{center}
\includegraphics[angle=90,width=.5\linewidth]{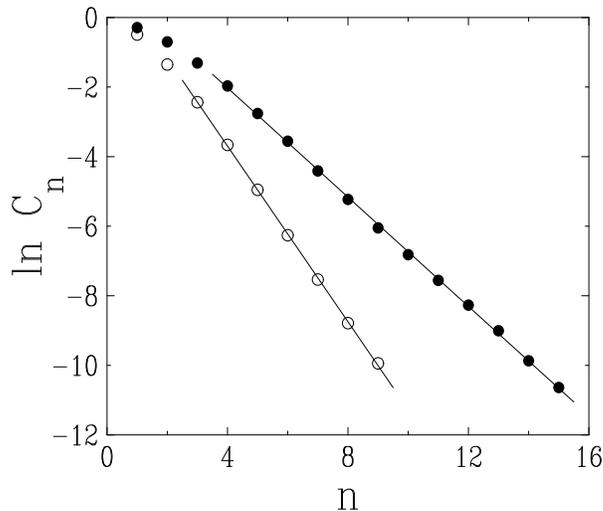}
\caption{\small
Logarithmic plot of the on-axis spin correlation function
$C_n$ against distance $n$.
Full symbols: standard dynamics on the honeycomb lattice.
Empty symbols: constrained dynamics on the square lattice.
Full straight lines: least-square fits with inverse slopes
$\xi_\he\approx1.27$ and $\xi_\ca\approx0.79$.}
\label{figcor}
\end{center}
\end{figure}

Figure~\ref{figcor}
shows a logarithmic plot of the on-axis spin correlation function,
against distance $n$.
The data at large enough distances are well fitted by straight lines,
demonstrating that correlations fall off exponentially to zero,
as $C_n\sim\exp(-n/\xi)$.
The inverse slopes of the least-square fits yield
$\xi_\he\approx1.27$ for the honeycomb lattice,
and $\xi_\ca\approx0.79$ for the square lattice.

\section{Discussion}
\label{dis}

The main focus of the present paper is on zero-temperature
single-spin-flip dynamics of ferromagnetic Ising models.
We have emphasized that there are two kinds of such dynamics.
The first kind ($\I$--type dynamics, in the classification of~\cite{ns})
generically corresponds to bona fide coarsening dynamics,
and lead to phase ordering by domain growth,
whereas the second kind ($\F$--type dynamics)
corresponds to descent dynamics,
and provides interesting examples of dynamical systems with many attractors.

The one-dimensional case has been reviewed in Section~\ref{chain}.
The existence of an exact mapping of the constrained zero-temperature dynamics
onto the dimer RSA problem opens up the possibility
of an exact analytical evaluation of many dynamical observables,
including characteristics of the attractors~\cite{usrsa}.
The comparison between the exact results thus obtained
and the predictions of the a priori ensemble
allows to test the validity of the so-called Edwards hypothesis,
in a regime which is very distant from the situation of gentle tapping,
considered originally by Edwards.
On the one hand, a flat average over the a priori ensemble of attractors
(especially in its refined version,
defined by imposing the value of the observed mean energy density)
provides inexact, but numerically reasonable,
predictions for physical quantities.
The numerical accuracy of a priori predictions may even by very good,
like e.g.~for the dynamical entropy (see Figure~\ref{figent}).
On the other hand, the energy correlation function provides an example
of a qualitative discrepancy between the actual behavior of attractors
and the prediction of any a priori ensemble.
Connected correlations indeed exhibit a factorial fall-off in $1/n!$
with distance, which is a generic characteristic of lattice RSA problems,
whereas any flat measure over an a priori ensemble
can be described by the transfer-matrix formalism,
and therefore leads to exponentially decaying correlations in one dimension.
Another example of a qualitative discrepancy concerns the lengths
of ordered clusters in one-dimensional spin models.
Cluster lengths are automatically statistically independent
and exponentially distributed in a priori ensembles,
again as a consequence of the underlying transfer-matrix formalism,
whereas both of these properties have been shown to be violated
in several examples of zero-temperature dynamics~\cite{bfs,uskawa}.

Examples of zero-temperature dynamics
on two-dimensional Ising models have then been investigated numerically.
The full dynamics on the honeycomb lattice
and the constrained one on the square lattice exhibit very similar features.
The observed logarithmic growth of the mean blocking time suggests that
the late stages of both examples of descent dynamics
are dominated by the relaxation of independent
two-level systems, in analogy with the one-dimensional situation.
This mechanism, in turn, is rather suggestive that attractors are likely
to have a non-trivial statistics.
The quantitative comparison with uniform measures on a priori ensembles
is far more difficult than in one dimension,
because a priori ensembles themselves may possess phase transitions
as a function of their density.
In particular, the accurately observed exponential fall-off
of spin correlations is not very informative in that respect.
Finally, it is worth noticing that the metastable states
put forward in a recent work on competitive cluster growth~\cite{lume}
share many of the features of the present two-dimensional attractors.

To sum up this investigation of the statistics of attractors
of zero-temperature descent dynamics in finite dimensions,
it turns out that
the so-called Edwards hypothesis does not hold in this context in general.
The predictions of the a priori approach, however,
often provide good numerical approximations.
This picture is corroborated by a recent work by Camia~\cite{camia}.


\Bibliography{99}

\bibitem{ange}
Angell C A 1995 Science {\bf 267} 1924

\bibitem{glassy}
Barrat J L, Dalibard D, Feigelman M, and Kurchan J eds 2003
{\it Slow Relaxations and Non\-equi\-li\-brium Dynamics in Condensed Matter}
Proceedings of Les Houches Summer School Session LXXVII (Berlin: Springer)

\bibitem{gold}
Goldstein M 1969 J. Chem. Phys. {\bf 51} 3728

\bibitem{tap}
Thouless D J, Anderson P W, and Palmer R G 1977 Phil. Mag. {\bf 35} 593

\bibitem{ks}
Kirkpatrick K and Sherrington D 1978 Phys. Rev. B {\bf 17} 4384

\bibitem{gs}
Gaveau B and Schulman L S 1998 J. Math. Phys. {\bf 39} 1517

\bibitem{bir}
Biroli G and Monasson R 2000 Europhys. Lett. {\bf 50} 155
\nonum
Biroli G and Kurchan J 2001 Phys. Rev. E {\bf 64} 016101

\bibitem{ldt}
Dean D S and Lef\`evre A 2001 Phys. Rev. Lett. {\bf 86} 5639
\nonum
Lef\`evre A and Dean D S 2001 J. Phys. A {\bf 34} L213
\nonum
Lef\`evre A 2002 J. Phys. A {\bf 35} 9037
\nonum
Dean D S and Lef\`evre A 2001 Phys. Rev. E {\bf 64} 046110

\bibitem{prbr}
Prados A and Brey J J 2001 J. Phys. A {\bf 34} L453

\bibitem{usrsa}
De Smedt G, Godr\`eche C, and Luck J M 2002 Eur. Phys. J. B {\bf 27} 363

\bibitem{gibbs}
Gibbs J W 1906 {\it The Scientific Papers of J Willard Gibbs}
(New York: Longmans, Green \& Co)
\nonum
Bumstead H A and Van Name R G eds 1961
{\it The Scientific Papers of J Willard Gibbs} in 2 vols (New York: Dover)

\bibitem{ll}
Landau L D and Lifshitz E M 1959 {\it Statistical Physics} (Oxford: Pergamon)
\nonum
Landau L D and Lifshitz E M 1960 {\it Electrodynamics of Continuous Media}
(Oxford: Pergamon)

\bibitem{fisher}
Fisher M E 1967 Physics {\bf 3} 255

\bibitem{langer}
Langer J S 1967 Ann. Phys. {\bf 41} 108

\bibitem{frenkel}
Frenkel J 1955 {\it Kinetic Theory of Liquids} (New York: Dover)

\bibitem{zinn}
Zinn-Justin J 1989 {\it Quantum Field Theory and Critical Phenomena}
(Oxford: Clarendon)

\bibitem{bustillos}
Ticona Bustillos A, Heermann D W, and Cordeiro C E 2004 J. Chem. Phys.
{\bf 121} 4804

\bibitem{sconf}
J\"ackle J 1981 Phil. Mag. B {\bf 44} 533
\nonum
Palmer R G 1982 Adv. Phys. {\bf 31} 669

\bibitem{ktw}
Kirkpatrick T R and Wolynes P G 1987 Phys. Rev. A {\bf 35} 3072
\nonum
Kirkpatrick T R and Thirumalai D 1987 Phys. Rev. B {\bf 36} 5388
\nonum
Kirkpatrick T R and Wolynes P G 1987 Phys. Rev. B {\bf 36} 8552
\nonum
Thirumalai D and Kirkpatrick T R 1988 Phys. Rev. B {\bf 38} 4881

\bibitem{sw}
Stillinger F H and Weber T A 1982 Phys. Rev. A {\bf 25} 978
\nonum
Stillinger F H and Weber T A 1984 Science {\bf 225} 983

\bibitem{fv}
Franz S and Virasoro M A 2000 J. Phys. A {\bf 33} 891

\bibitem{bm}
Berg J and Mehta A 2001 Europhys. Lett. {\bf 56} 784
\nonum
Berg J and Mehta A 2001 Adv. Complex Syst. {\bf 4} 309

\bibitem{pbt}
Prados A and Brey J J 2001 Phys. Rev. E {\bf 66} 041308

\bibitem{edwards}
Edwards S F 1994 in {\it Granular Matter: An Interdisciplinary Approach}
Mehta A ed (New York: Springer)

\bibitem{fdt}
Cugliandolo L F and Kurchan J 1994 J. Phys. A {\bf 27} 5749
\nonum
Cugliandolo L F, Kurchan J, and Parisi G 1994 J. Phys. I (France) {\bf 4} 1641
\nonum
Cugliandolo L F, Kurchan J, and Peliti L 1997 Phys. Rev. E {\bf 55} 3898

\bibitem{ba}
Barrat A, Kurchan J, Loreto V, and Sellitto M 2001 Phys. Rev. Lett. {\bf 85}
5034
\nonum
Barrat A, Kurchan J, Loreto V, and Sellitto M 2001 Phys. Rev. E {\bf 63} 051301

\bibitem{red}
Spirin V, Krapivsky P L, and Redner S 2001 Phys. Rev. E {\bf 63} 036118
\nonum
Spirin V, Krapivsky P L, and Redner S 2002 Phys. Rev. E {\bf 65} 016119

\bibitem{ns}
Nanda S, Newman C M, and Stein D L 2000 in {\it On Dobrushin's Way
(from Probability Theory to Statistical Physics)} Dubrushin R L \etal eds
Amer. Math. Soc. Transl. {\bf 198}
\nonum
Newman C M and Stein D L 2000 Physica A {\bf 279} 159
\nonum
Nanda S, Newman C M, and Stein D L 2004 preprint cond-mat/0411286

\bibitem{barc}
Sollich P and Ritort F eds 2002 {\it Workshop on glassy behaviour
of kinetically constrained models} J.~Phys. Cond. Matt. {\bf 14} 1381-1696

\bibitem{riso}
Ritort F and Sollich P 2003 Adv. Phys. {\bf 52} 219

\bibitem{glau}
Glauber R J 1963 J. Math. Phys. {\bf 4} 294

\bibitem{bray}
Bray A J 1994 Adv. Phys. {\bf 43} 357

\bibitem{flory}
Flory J P 1939 J. Am. Chem. Soc. {\bf 61} 1518

\bibitem{core}
Cohen E R and Reiss H 1963 J. Chem. Phys. {\bf 38} 680

\bibitem{rsa}
Evans J W 1993 Rev. Mod. Phys. {\bf 65} 1281

\bibitem{gumbel}
Gumbel E J 1958 {\it Statistics of Extremes} (New York:
Columbia University Press).

\bibitem{cri}
Crisanti A, Ritort F, Rocco A, and Sellitto M 2000 J. Chem. Phys. {\bf 113}
10615

\bibitem{bfs}
Berg J, Franz S, and Sellitto M 2002 Eur. Phys. J. B {\bf 26} 349

\bibitem{baxter}
Baxter R J 1982 {\it Exactly Solved Models in Statistical Mechanics}
(London: Academic)

\bibitem{tm}
Crisanti A, Paladin G, and Vulpiani A 1992
{\it Products of Random Matrices in Statistical Physics}
(Springer, Berlin, 1992)
\nonum
Luck J M 1992 {\it Syst\`emes d\'esordonn\'es unidimensionnels}
in French (Saclay: Collection Al\'ea)

\bibitem{uskawa}
De Smedt G, Godr\`eche C, and Luck J M 2003 Eur. Phys. J. B {\bf 32} 215

\bibitem{lume}
Luck J M and Mehta A 2004 preprint cond-mat/0410385

\bibitem{camia}
Camia F 2004 preprint cond-mat/0410543

\endbib
\end{document}